# Dynamics of osmotic flows

Patrice BACCHIN[a,*]

[a]Laboratoire de Génie Chimique, Université de Toulouse, CNRS, INPT, UPS, Toulouse, France
*corresponding author: patrice.bacchin@univ-tlse3.fr

The paper presents a theoretical model that allows the dynamic description of osmotic flows through a semi-permeable interface. To depict the out-of-equilibrium transfer, the interface is represented by an energy barrier that colloids have to overcome to be transmitted to the other side of the membrane. This energy barrier thus represents the selectivity of the membrane. Furthermore, this energy barrier induces additional force terms in the momentum and the mass balances on the fluid and the colloids phases. Based on a two- fluid model, these forces can reproduce the physics of the osmotic flow without the use of the semi-empirical laws of non-equilibrium thermodynamics. It is shown that a decrease in local pressure near the interface initiates osmosis. When these balance equations are solved in a transient mode, the dynamic of the osmotic flow can be described. The paper illustrates these potentialities by showing the dynamic of an osmosis process occurring in the absence of transmembrane pressure and both the dynamic of the reverse osmosis with a constant flow through the membrane. A simulation reproducing the dynamic of the Abbé Nollet experiment is presented. The role played by the colloid-membrane interactions on the osmotic flow mechanism and on the counter osmotic pressure is analyzed and discussed in great details.

## 1 Introduction

The understanding of the transport of colloids at, or across, interfaces is still a scientific challenge meeting applications in many processes. For example, flow through semi-permeable membranes is a common process in living bodies (kidneys, membrane cells, etc.) and in industrial applications (filtration, desalting, etc.). Beyond these applications, the recent development of microfluidic experiments and of nano-scale engineered interfaces has revived the question of the role played by colloid-surface interactions on the transport at, or across, interfaces[1]. When considering the colloid transport across interfaces, the classical model for osmotic flow derives from the semi-empirical formulation of Kedem and Katchalsky[2] that considers non-equilibrium thermodynamics with the assumption of linearity between the fluxes and the driving forces. The mechanical approach (adapted from Darcy law) and the thermodynamic formulation converges to the writing of the velocity of the solvent, $u_w$, and the colloids, $u_c$, through the membrane [3] :

$$u_w = -\frac{k_w}{\eta}\left(\frac{dp}{dz} - \sigma\frac{d\Pi_{cc}}{dz}\right) \quad (1)$$

$$u_c = (1-\sigma)u_w - P\frac{dc}{dz} \quad (2)$$

Where p is the pressure, $\Pi_{cc}$ is the osmotic pressure due to the colloid-colloid interactions, $\sigma$, $k_w$ and P are respectively the reflection coefficient, water and the solute permeabilities. In this model, the membrane is considered as a discrete transition region (an active layer assumed to be completely uniform across its thickness) between two homogeneous solutions. It is therefore assumed that flows through the membrane are caused by the differences in potentials occurring across the membrane.

These equations describe the fluxes between two compartments by assuming that the drive is the difference in chemical potentials. However, one has to note that the effective forces leading to the transport are not accounted by this formalism. Early works on osmosis underline the importance played by the interaction between colloids or molecules and membrane interface on the osmosis flows. Van't Hoff explained osmosis in terms of the work done by the rebounding molecules of a solute on a selective semipermeable membrane: "The mechanism by which, according to our present conceptions, the elastic pressure of gases is produced is essentially the same as that which gives rise to osmotic pressure in solutions. It depends, in the first case, upon the impact of the gas molecules against the wall of the vessel; in the latter, upon the impact of the molecules of the dissolved substance against the semipermeable membrane, since the molecules of the solvent, being present upon both sides of the membrane through which they pass, do not enter into consideration [4]". Einstein [5] considered that the colloids exerted a pressure on the material at the origin of the partition: "We must assume that the suspended particles perform an irregular move (even if a very slow one) in the liquid, on account of the molecular movement of the liquid; if they are prevented from leaving the volume V* by the partition, they will exert a pressure on the partition just like molecules in solution". Fermi [6] stated that the pressure on the side of the membrane facing the solution is increased by the impacts of the molecules of the dissolved substances, which cannot pass through the membrane.

This interest for considering the effects of interactions with the membrane has been recently discussed [7–10] but there is still little knowledge on how the colloid-interface interactions play a role on the dynamic of osmotic flow and how these interactions are related to the properties of the fluid and the membrane. The aim of this paper is to put forward a model that implements the role of the colloid-membrane interactions on the dynamics of osmotic flow.

# 2 Theoretical background

A new model has been recently established from the momentum balance for the fluid and the colloid phase on an energy landscape [11]. The concept of energy landscapes [12] allows the mapping of the colloid-membrane interaction energy (related to the Gibbs free energy that can also be expressed per unit of volume as a pressure, $\Pi_i$ named interfacial pressure in the paper) for all of the spatial positions of the colloids in the vicinity or inside the membrane. This map represents the overall interactions between the colloids and the membrane interface (as for example, DLVO and hydration forces [13–15]) but can also account for the energetic changes in the colloid conformation required for the transport in a spatial direction (for example, for deformable particles, or for extensible or unfolding proteins) [16]. The next sections are establishing the TFEL Two-Fluid model on an Energy Landscape (section 2.1) and will compare the model with the approach derived from the Non Equilibrium Thermodynamic (NET) approach (section 2.2). The set of equations to solve transient osmotic flow in one direction in a dimensional is presented in section 2.3 and its non-dimensional form in section 2.4.

## 2.1 Two-Fluid model on an Energy Landscape (TFEL)

In the two-fluid (or mixture) model [17,18], the momentum balances are established onto an energy landscape with the two-fluid model formalism i.e. for the fluid phase, the colloid phase having a velocity, $u_c$ (Eq. 3) and, by addition of the balance on these two phases, on the mixture phase having a velocity, $u_m$ (Eq. 4):

Momentum balance
On the dispersed phase

$$+ nF_{drag} - \nabla \Pi_{cc} - \phi \nabla \Pi_i = 0 \qquad (3)$$

On the fluid

$$-\frac{\eta_m u_m}{k_p} - nF_{drag} - \nabla p + \eta_m \nabla^2 u_m + \nabla \Pi_{cc} = 0 \qquad (4)$$

On the mixture

$$-\frac{\eta_m u_m}{k_p} - \nabla p + \eta_m \nabla^2 u_m - \phi \nabla \Pi_i = 0 \qquad (5)$$

Mass balance
On the dispersed phase

$$\frac{\partial \phi}{\partial t} = -\nabla \cdot (\phi u_c) \qquad (6)$$

On the fluid

$$\frac{\partial(1-\phi)}{\partial t} = -\nabla \cdot \left((1-\phi)\boldsymbol{u_f}\right) \quad (7)$$

On the mixture

$$0 = \nabla \cdot \boldsymbol{u_m} \quad (8)$$

The different contributions in these equations are dissipative or elastic in nature. The dissipative contributions are:

- the drag force which represents the forces due to the friction induced by the relative velocity between the phases (**colloid-fluid** friction) and the colloid mobility, m :

$$\boldsymbol{F}_{drag} = \frac{u_m - u_c}{m(\phi)} \quad (9)$$

- the viscous dissipation, $-\eta_m \nabla^2 \boldsymbol{u_m}$, due to the viscosity mixture, $\eta_m$, in Eqs. 4 and 5 (**fluid-fluid** friction induced by the shear)
- the viscous dissipation in the fluid due to the interface that can be linked to a porous media permeability, $-\frac{\eta_m}{k_p}\boldsymbol{u_m}$ in Eqs. 4 and 5 (**fluid-membrane** friction) where, $k_p$, is the permeability coefficient of the membrane.

The elastic storage (non-dissipative contributions) are:

- the thermodynamic (reversible) colloid pressure gradient, $\nabla \Pi_{cc}$, that corresponds to the water activity difference (**colloid-colloid** interaction)
- the interfacial pressure, $\phi \nabla \Pi_i$, in Eqs. 3 and 5 (**colloid-membrane** interaction)
- the pressure drop, $-\nabla p$, representing the energy dissipated in the system (**fluid-fluid** interaction)

These equations can be solved in 2D or 3D in order to analyze the influence of the colloid-wall interactions on the transport of the fluid and of the colloids. However, this paper focuses on solving the equations in 1D and in the absence of shear to keep the system as simple as possible and to estimate the potentialities of the essential ingredient of the model: the energy landscape mapped with the interfacial pressure, $\Pi_i$. In the next section, the model is compared to the classical Kedem Katchlasky approach developed from a NET approach.

## 2.2 Comparison of TFEL model with the NET approach

In 1D and in a steady state, the continuity equation for the fluid (Eq. 8) results in considering that the mixture velocity, $u_m$, is constant along the distance. Eqs. 3-9 can be combined to define the mixture and the colloids' velocities:

$$\frac{\phi}{V_p}\frac{u_m - u_c}{m(\phi)} - \nabla \Pi_{cc} - \phi \nabla \Pi_i = 0 \quad (10)$$

$$\frac{\eta_m}{k_p} u_m + \nabla p + \phi \nabla \Pi_i = 0 \quad (11)$$

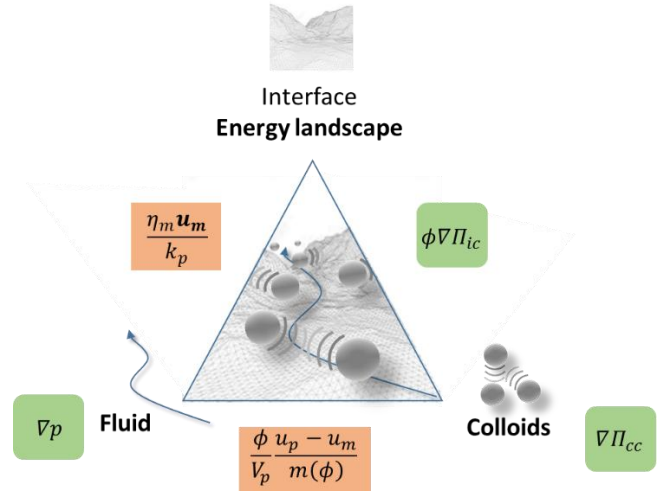

*Fig. 1 : Schematized representation of the different terms playing a role in momentum balances (Eqs. 10 and 11) on a ternary diagram with the two-fluid phases (the fluid and the dispersed phase (the colloids)) and the interface characterized by the interfacial pressure map. The different terms of the equations are placed in the figure according to the coupling interactions between these different phases. Some of these terms are dissipative (in orange rectangles) and some are elastic (in green rounded squares).*

The momentum balance on colloids (Eq. 10 resulting from Eqs. 3 and 9) establishes that the drag force acting on colloids because of the slip velocity (first term of Eq. 10) is the result of the osmotic pressure gradient (second term) and the interfacial pressure gradient due to colloid-membrane interactions (third term). In the momentum mixture balance (Eq. 11 resulting from Eq. 5), the pressure drop due to friction between the mixture and the interface (first term of Eq. 4) is brought by the fluid pressure gradient (second term) and by the colloid membrane interaction. The colloid membrane interactions play here the role of a forcing term on the momentum equations of the fluid flow, similarly to a Force Coupling Method[19]. This mathematical writing therefore enables accounting for the interactions between the different phases as schematized in Fig. 1. Eq.10 indicates that the drag force acting on particles (dissipation due to the **fluid-colloid** slip velocity) is stored in colloid-colloid interactions (through the osmotic pressure, $\Pi_{cc}(\phi)$) or in colloid-membrane interactions (through the interfacial pressure, $\Pi_i(x)$). Eq. 11 means that the drag forces on the interface (**fluid-membrane** dissipation) are induced by the destocking of the fluid pressure (**fluid-fluid** interactions) or **colloid-membrane** interactions. Eqs. 10 and 11 allow to express the mixture velocity and the colloid velocity:

$$\boldsymbol{u}_m = -\frac{k_p(z)}{\eta}\left(\frac{dP(z)}{dz} + \phi\frac{d\Pi_i(z)}{dx}\right) \quad (12)$$

$$\boldsymbol{u}_c = \boldsymbol{u}_m - \frac{V_p}{\phi}m(\phi)\left(\frac{d\Pi_c(\phi)}{dz} + \phi\frac{d\Pi_i(z)}{dz}\right) \quad (13)$$

These last two equations can be compared to Eqs. 1 and 2 resulting from the non-equilibrium KK approach [2]. In the equation describing the fluid or the mixture flow through the membrane, the forced convection term (due to the gradient of pressure, $\frac{dP(z)}{dz}$) is balanced by the osmotic flow which is written, $-\sigma\frac{d\Pi_{cc}}{dz}$, with the NET approach (Eq. 1) and by the interfacial pressure, $\phi\frac{d\Pi_i(z)}{dz}$, with the Two Fluid on Energy Landscape TFEL (Eq. 12). In the NET approach the osmotic flow is a function of the difference in osmotic pressure (also linked to the difference in water activity and then the water chemical potential difference) whereas, in the TFEL model the osmotic flow is seen as the consequence of the colloid-membrane interaction. When these equations are integrated, the resulting counter osmotic pressure is $-\sigma\Delta\Pi_{cc}$ and $\int \phi d\Pi_i$ respectively. The counter pressure is therefore not the direct result of the concentration gradient, but is rather depicted as the consequence of the exclusion of particles by the colloid-interface interactions. The counter pressure only occurs in the zone where an energy map gradient is present (referred to as exclusion layers in this paper). The TFEL approach (based on a mechanical approach) is closely related to the thermodynamic approach when a potential-energy profile is introduced in KK equations [20] to describe the osmotic transport across a membrane.

The colloid-membrane interaction is closely linked to the osmotic pressure difference in KK approach (Eq. 1). When considering the momentum balance on the colloids (eq. 10), the drag force on the colloids is balanced by the collective diffusion (osmotic pressure gradient) and the migration of colloids near the interface (interfacial pressure gradient).

$$\phi\frac{d\Pi_i(x)}{dz} = -\frac{d\Pi_{cc}(\phi)}{dz} + \frac{\phi}{V_p}\frac{\boldsymbol{u}_m - \boldsymbol{u}_c}{m(\phi)} \quad (14)$$

However, colloid-interface interactions and colloid-colloid interactions are linked. As expressed by the integration of Eq. 3 (or Eq. 13), these colloid-interface interactions are also balanced by the osmotic pressure gradient and the drag forces:

$$\int_{Ex}\phi d\Pi_i = -\int_{Ex} d\Pi_{cc} + \int_{Ex} n\boldsymbol{F}_{drag}dx \quad (15)$$

Physically, the counter pressure is then due to the osmotic pressure difference at the exclusion layer boundaries (it can be seen as the osmotic flow contribution: a counter osmotic pressure) and to the drag force acting on the particles (a counter drag pressure).

$$\boldsymbol{u}_m = \frac{L_p}{\eta}\left(\Delta P + \int_{Ex} d\Pi_{cc} - \int_{Ex} n\boldsymbol{F}_{drag}dx\right) \quad (16)$$

If considering a very thin exclusion layer, the integral of the drag force in the exclusion layer is negligible, $\int_{Ex} n\boldsymbol{F}_{drag}dx = 0$. In such conditions (details are given in supplementary information 1), the osmotic pressure difference is related to the concentration difference (if considering an ideal gas of colloids) and to the partition coefficient, $\Phi$, that is classically used to defined the "exclusion" role of a membrane:

$$\int_{Ex} d\Pi_{cc} = -(1 - \Phi)\Delta\Pi_{cc} \quad (17)$$

In these conditions, the TFEL model can be linked to the KK approach by considering that the Staverman coefficient is:

$$\sigma = 1 - \Phi \qquad (18)$$

The limit for the counter pressure is then similar to the osmotic pressure difference with a Staverman coefficient [21] defined as $\sigma = (1 - \Phi)$ where $\Phi$ is the partition coefficient. This result is coherent with the frequent writing of the Staverman coefficient as a reflection coefficient accounting for the leakage of a membrane ($\sigma$ ranging from 0 for a completely non-retentive membrane to 1 for a membrane impermeable to the solute). Finally, the model proposes an analytical writing for the Staverman coefficient which is still the subject of open questions and is differently interpreted from the Kedem and Katchalsky approach [22].

### 2.3 Set of equations and data for the dynamic description

The TFEL model can be used to depict the dynamic of the osmotic flow. To illustrate this ability of the model, the transfer of a solute is considered along the pore axis or through the dense membrane thickness. The problem is treated in 1D (the z direction normal to the membrane surface) with no shear ($\eta_m(\phi)\dot{\gamma} = 0$), and when the colloids pressure, $\Pi_{cc}$, has only a thermodynamical contribution (i.e. in the absence of deposit or gel on the membrane surface). Eqs. 10 and 11 can be written as a set of partial differential equations along the z direction :

$$-\frac{\eta_m}{k_p}u_m \quad -\frac{dp}{dz} \quad -\phi\frac{d\Pi_i(z)}{dz} = 0 \qquad (19)$$

$$\frac{\partial \phi}{\partial t} = -u_m \frac{d\phi}{dz} - \frac{d\left(m(\phi)V_p\left(-\frac{d\Pi_{cc}(\phi)}{dz} - \phi\frac{d\Pi_i(z)}{dz}\right)\right)}{dz} \qquad (20)$$

Where $u_m$ represents the permeate flux through the membrane being constant along z from continuity consideration (eq. 8). The interfacial pressure along z is defined with a function (represented in Fig. 2) having a continuous first order derivative (details on SI 2): the derivative of the interfacial pressure in Eqs 19 and 20 is therefore a continuous analytical function. The two main parameters for this function are the height of the interfacial pressure (the maximum value that represents the height of the energy map) and the range of the interfacial pressure (that first represents the thickness of the exclusion layer and then the stiffness of the energy peak). The height or the maximum of the interfacial pressure is taken at $\frac{V_p\Pi}{kT} = 2.3$ that should correspond, for infinitively thin exclusion layer, to a partition coefficient, $\Phi = e^{-\frac{V_p\Pi}{kT}}$, of 0.1 (according to the analogy between the maximum interfacial pressure and the partition coefficient in SI 3).

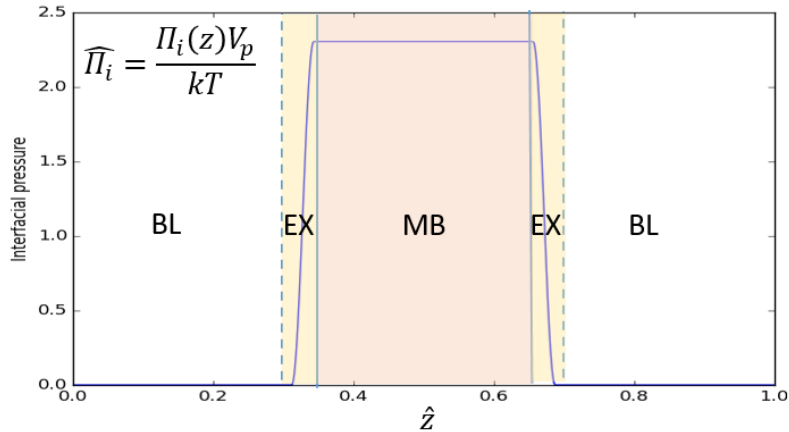

*Fig. 2 : Representation of the energy map describing the solute-membrane interactions through the interfacial pressure. The membrane is represented with two exclusion layers, EX (where the variations of the interfacial pressure are localized) and the core of the membrane, MB, (where the interfacial pressure is maximum). The main parameters are the maximum value of the interfacial pressure (that is related to the overall membrane selectivity) and the exclusion layer thickness (that is related to the stiffness of solute-membrane interactions). The energy landscape is surrounded by two boundary layers or polarization layers, BL, allowing the description of the interfacial mass transfer.*

## 2.4 Dimensionless set of equations

In the results section, the results will be presented in a non-dimensional form in order to generalize the application of the model. Eqs. 19 and 20 can be modified as :

$$-\frac{\eta_m}{\widehat{k_p}} \boldsymbol{Pe} - \nabla \hat{p} - \phi \nabla \widehat{\Pi_l} = 0 \quad (21)$$

$$\frac{\partial \phi}{\partial \hat{t}} = -\nabla \cdot \left( \phi (Pe - K(\phi) \nabla \widehat{\Pi_l}) - K(\phi) \nabla \widehat{\Pi_{cc}} \right) \quad (22)$$

where the different terms are made dimensionless by dividing the initial equations by the diffusion characteristic parameter. The dimensionless quantities are given in Table 1 where, $a$, is the size of the solute and, $\delta$, is the characteristic dimension of the system (will be taken as the total thickness of the system).

The set of equations has been solved with the partial differential equation solver Fipy [23] implemented on the Python platform Canopy (Enthought, Austin). The full code used for this solving is given in SI 4. In order to use the formalism of the fipy solver, the convection term is written as the combination of the Péclet number and a Péclet number due to the migration velocity induced by the presence of the interface $Pe_i = K(\phi) \nabla \widehat{\Pi_l}$. This last term is proportional to the gradient of the interface pressure (the slope of the energy map). The dimensionless diffusive term is given by the generalized Stokes Einstein law, $K(\phi) \nabla \widehat{\Pi_{cc}}$.

Table 1 : The dimensionless quantities used to define the dynamic osmotic problem. The correspondence with the dimensional quantities are given for the conditions of $a=10^{-8}$ m, $\delta=10^{-6}$ m, $\mu=10^{-3}$ Pa.s and, $T=298$ K.

| Quantity | Dimensionless writing | Correspondence |
|---|---|---|
| Velocity | Péclet $Pe = \frac{u_m \delta}{m_0 kT}$ | $u\ (m) = 2.18\ 10^{-5}\ Pe$ |
| Permeability | $\widehat{k_p} = k_p \frac{9}{2a^2}$ | $k_p(m^2) = 0.222\ 10^{-16} \widehat{k_p}$ |
| Pressure | $\hat{p} = \frac{V_p p}{kT}$ | $p(Pa) = 982\ \hat{p}$ |
| Time | $\hat{t} = \frac{m_0 kT}{\delta^2} t$ | $t(s) = 0.0458\ \hat{t}$ |
| Osmotic pressure or interfacial pressure | $\widehat{\Pi} = \frac{V_p \Pi}{kT}$ | $\Pi(Pa) = 982\ \widehat{\Pi}$ |
| Mobility | Hindrance settling coefficient $K(\phi) = \frac{m}{m_0}$ | $m\ (kg^{-1}.s) = 5.31\ 10^9\ K(\phi)$ |

# 3 Transient description of osmotic flows

The ability of the model to describe the dynamics of the osmotic flow is illustrated for two different and complementary case studies (Fig. 3):

- The Reverse Osmosis at a constant flow rate: the model dynamically describes the local concentration through the membrane, the selectivity of the membrane and the increase of the counter pressure

- The 1748 Abbé Nollet osmosis experiment [24] where two compartments with two different concentrations induce an osmotic flow that change the height of the compartment and then the static pressure across the membrane. The model describes the transient change in height and the equilibrium reached when the difference in pressure (due to the difference in liquid level) is compensating the osmotic pressure.

The main conditions that define these two cases are illustrated in Fig. 3.

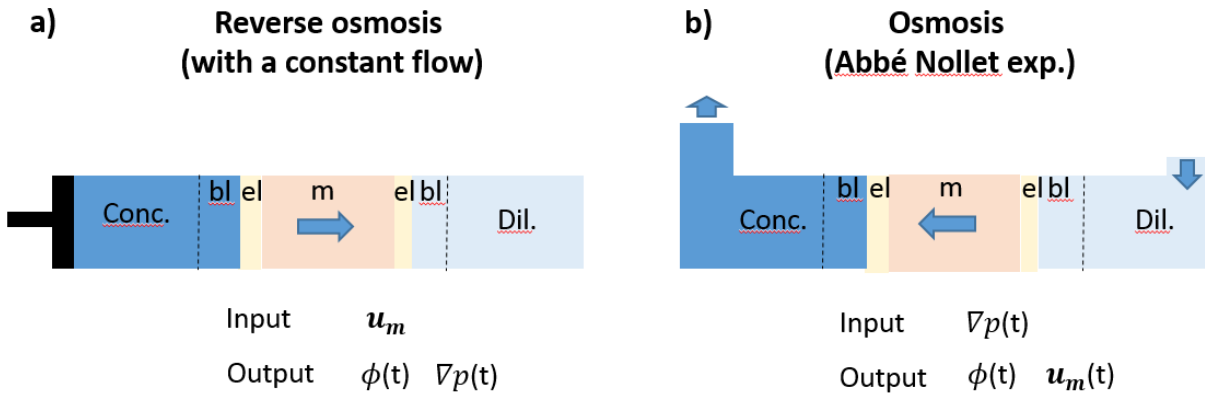

*Fig. 3 : Representation of the conditions for the two case studies a) reverse osmosis with constant flow b) Abbé Nollet osmosis experiment. The simulations are performed through the membrane system composed of the boundary layers (bl), the exclusion layers (el) and the membrane (m). The reverse osmosis case is simulated by considering a fixed flow rate from the concentrate (conc.) towards the diluate (dil.) compartment; the model then describes the subsequent transient increase in transmembrane pressure. For the case b), according to the conditions of the original Abbé Nollet experiment, the simulation is performed by considering that the transmembrane pressure is imposed by the difference of the height of the liquid between the concentrate and the diluate compartments. The simulation allows the calculation of the osmotic flow and, from a mass balance on the concentrate and diluate compartments, the evolution of the liquid level (that subsequently modifies the transmembrane pressure).*

For the reverse osmosis simulation (Fig. 3a), the boundary conditions for the volume fraction are a specific concentration on the external side of the concentrate boundary layer and an outlet flow condition (gradient of volume fraction equated to zero) at the external of the diluate boundary layer. The flow rate through the membrane, $u_m$, being given, solving of Eq. 20 gives the transient evolution of the concentration profile. The simulation first enables the concentration coming in the diluate side and then the transient reverse osmosis selectivity to be determined. Once solved, the concentration profile helps the counter pressure to be determined by integration of Eq. 19. Such solving, corresponding to a constant flux filtration mode, will be presented in section 3.1.

To resolve a constant pressure mode (Fig. 3b), Eqs. 19 and 20 have to be solved simultaneously. This is for example the case of the Abbé Nollet experiment for osmosis where the pressure in the compartments is fixed by the liquid heights. The boundary conditions for the concentration at the external side of the boundary layers are also fixed by the experiment conditions. From these boundary conditions, the simulations allow the transient variation of the concentration profile and the osmotic flux to be determined. As a consequence of the osmotic flow, the volume of the liquid in the two compartments changes. The changing heights of the compartments thus induce a change in the fluid pressure acting on the membrane. The change in volume also modifies the solute concentration in the compartment (this last effect can be neglected if the volume of the compartment is large enough). The full set of equations (given in SI 5) complete the equations that describes the transfer of the fluid and the solute (Eqs. 19 and 20) with global and partial mass balance equations for the concentrate and the diluate compartments.

### 3.1 Reverse osmosis with constant flow

The reverse osmosis simulations are performed for a Péclet number of 2. The variation of the volume fraction profile with time is given in Fig. 4. The related variations of the solute transmission and the counter osmotic pressure with time are given in Fig. 5 a) and b) respectively. In these simulations, the initial volume fraction is zero all along z. At t=0, the volume fraction is 0.001 in the concentrate side (left side in Fig. 4). In first simulation time, the solute is then diffusing toward the membrane. When the solute reaches the exclusion layer, the solute is excluded by the membrane (due to the gradient of the interfacial pressure in Fig. 2). The counter pressure that was initially zero starts to increase. For longer time, the exclusion leads to the solute accumulation in the concentrate side and, consequently, to an increase in the solute transmission. The counter pressure reaches a slight maximum at a dimensionless time around 0.2. In parallel, the transmission progressively increases to converge towards a stationary value. As already discussed in a previous paper [11], the variation of the steady state transmission with the Péclet closely follows the relationships obtained by considering a partition coefficient and therefore a concentration jump instead of the exclusion layers. At a steady state, one can note that the volume fraction difference is $1.3 \cdot 10^{-3}$ (Fig. 4) whereas the dimensionless counter pressure is $1.2 \cdot 10^{-3}$ (fig. 5). The counter pressure is very close to $(1 - \Phi)\Delta\Pi_{cc}$ where $\Phi$ is the partition coefficient (0.1 as discussed in section 2.3).

More precisely, the different contribution to the counter pressure, CP, are decomposed as follows (from Eq. 15) :

$$CP = -\int_{Ex1} d\Pi_{cc} - \int_{Ex2} d\Pi_{cc} + \int_{Ex} n\mathbf{F}_{drag} dx \quad (23)$$

The numerical values for the different contributions are given below :

$$1.2 \cdot 10^{-3} = -(0.165 - 1.6)10^{-3} - (0.3 - 0.038)10^{-3} + 0.027 \cdot 10^{-3} \quad (24)$$

The counter pressure is thus mainly due to the difference in osmotic pressure in the exclusion layers. The drag force contribution to the counter pressure represents here around 2.2 % of the total counter pressure.

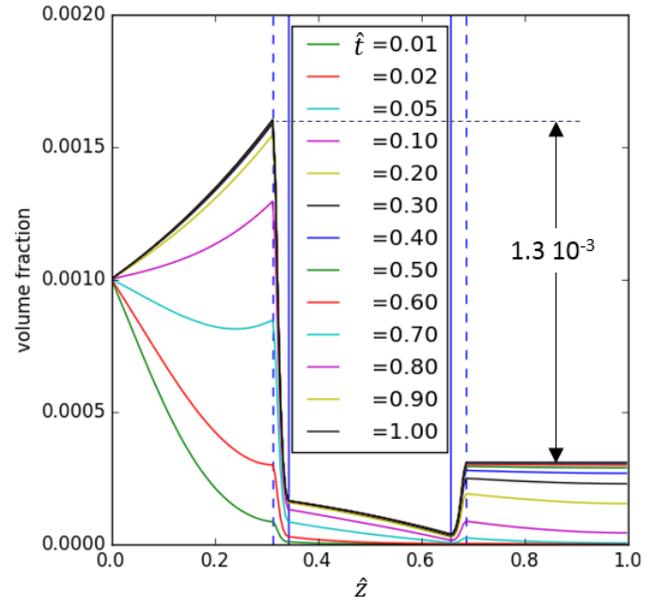

*Fig. 4 : Time variations of the profiles of colloids volume fraction. The volume fraction is plotted as a function of the dimensionless distance, $\hat{z}$, through the boundary layer, the exclusion layers (between the dashed and full vertical lines) and the membrane (between the full vertical lines) for different dimensionless time, $\hat{t}$.*

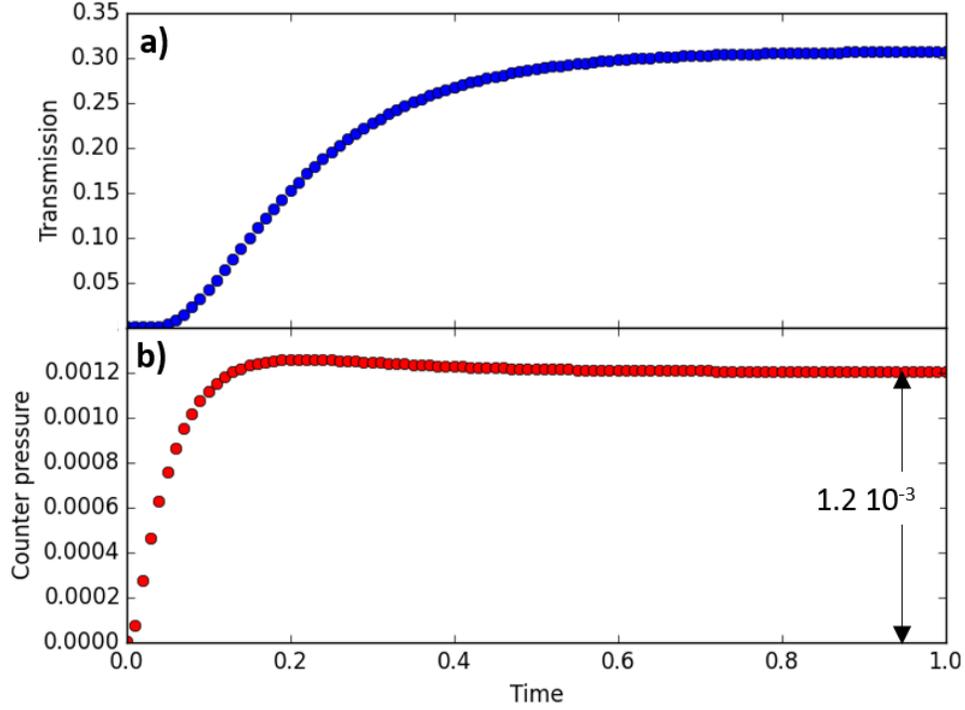

*Fig 5 : a) Variation of the colloids transmission defined as the ratio of the concentration in the dilute and the concentrate side b) Variation of the dimensionless counter pressure as a function of the dimensionless time.*

### 3.2 Abbé Nollet osmosis experiments

For the osmosis simulation, the volume fractions are fixed on the external side of the boundary layers. Initially, the solute concentration in the different layers is zero and the volume fraction is 0.1 on the boundary on the concentrate compartment; such conditions could mimic a rapid introduction of the solute in the concentrate compartment. The solute is therefore diffusing from the concentrate to the dilute side; the corresponding concentration profiles are presented in Fig. 6. The interactions of the solute with the membrane (defined by the interfacial pressure function defined in section 2.3) prevent the diffusion of the solute inside the membrane. This exclusion induces the osmotic flow (Fig. 7b) that is defined by a negative Péclet value (the flow is from the diluate to the concentrate side i.e. in the direction of decreasing z). The omostic flow induces an increase in the liquid level in the concentrate side and an opposite decrease of the level in the dilute side. These levels are translated into a resulting static pressure in Fig. 7a. The difference in the static pressure therefore induces a forced advection opposite to the osmotic flow. The equilibrium is reached when the difference in static pressure, 0.0754, compensates the part of the difference of the osmotic pressure corresponding to the external boundary of the exclusion layers. In the end, the contributions to the difference in pressure across the membrane, $\Delta \hat{p}$, are given according to Eq. 15:

$$\Delta \hat{p} = - \int_{Ex1} d\Pi_{cc} - \int_{Ex2} d\Pi_{cc} + \int_{Ex} n\boldsymbol{F}_{drag} dx \quad (25)$$

$$0.0754 = -(0.0088 - 0.0905) - (0.0071 - 0.0011) - 0.0003 \quad (26)$$

The difference of pressure is still close to the difference in osmotic pressure (0.0834 as indicated in Fig. 5) weighted by the partition coefficient according to: $(1 - \Phi)\Delta\Pi_{cc} = (1 - 0.1)0.0834 = 0.075$. There is, in this case, a residual drag force that is very low and negative. The low value is due to the fact that the flow is very low in conditions near to the equilibrium, and the negative value results from the negative osmotic velocity (directed from the diluate toward the concentrate compartment).

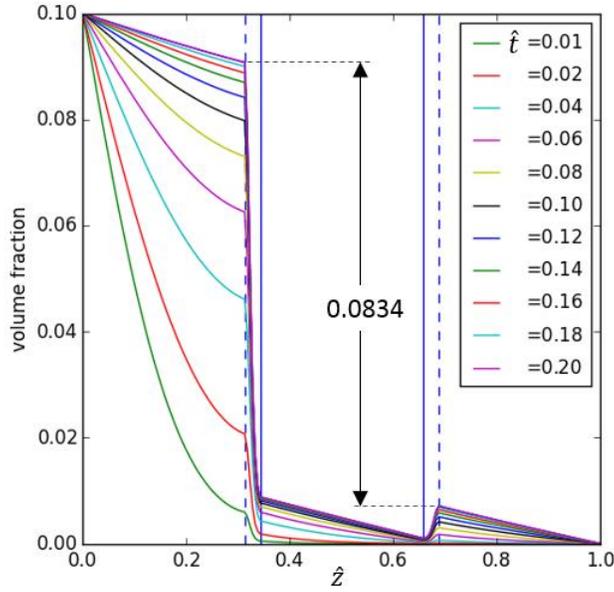

*Figure 6 : The dynamic variation of the volume fraction through the membrane during osmosis. At equilibrium, the difference of the volume fraction at the exclusion layers boundary is 0.0834.*

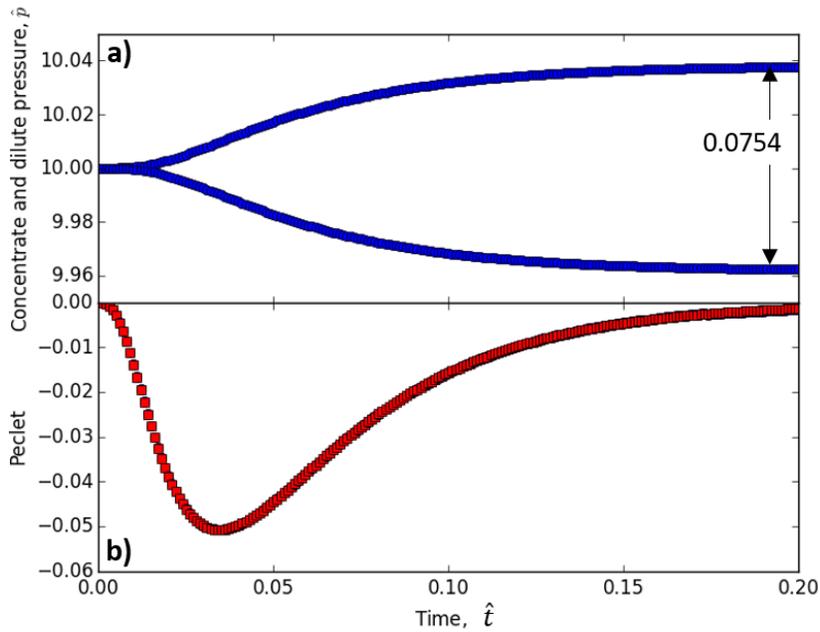

*Figure 7 : a) The variation of the pressure in the concentrate and dilute side due to the liquid level changes induced by the osmotic flow. b) Variation of the Péclet number (that represents the osmotic flow) with the dimensionless time. The negative value of the Péclet indicates that the osmotic flow is directed towards the concentrate compartment (direction of decreasing z). The osmotic flow is maximum when the concentration difference is great and as soon as the difference in pressure is not too significant. For longer time, the osmotic flow tends to zero; the equilibrium is reached when the difference in dimensionless pressure generating a forced convection compensates the osmotic flow due to the difference in the volume fraction.*

# 4 Discussion

The Two Fluids on Energy Landscape (TFEL) model allows the description of the transfer of solute having an interaction with its surrounding environment. The model unravels the effect of the combination of interactions between the three bodies (Fig. 1): the fluid, the solute and its environment (a membrane interface in this paper). In the case of the flow of a mixture of fluid and colloid through a membrane, the model describes the flow without applying the Kedem-Katchalsky approach as summarized in section 4.1. In the end, the fluid and the colloid flows can be represented as the result of driving pressures that are the combination of the fluid, the interfacial and the osmotic pressures (section 4.2). From the momentum balances implemented in the model, it can be demonstrated that the interfacial pressure leads to a local decrease in the fluid pressure (section 4.3) that participates with the osmotic flow mechanism discussed in section 4.4. Such a phenomena can be explained by the irregular movement of colloids in the exclusion layers near the membrane and by a local violation of Newton's third law for the mixture (section 4.5). The model also brings new insight into the controversy about the existence of a pressure drop in polarization layers during filtration (section 4.6).

## 4.1 Ability of the TFEL model to describe osmotic membrane transport

The model provides a continuous description in one shot of the transfer through an interface without introducing boundary conditions at the membrane wall. This potentiality is mainly due to the description of the transport on an energy landscape that represents the interaction between a solute and the membrane in their large diversity [11]. Presence of the membrane is then accounted by the solute-membrane interactions (the Energy Landscape part of TFEL) that described the solute exclusion at the origin of the membrane selectivity. It is therefore not necessary to introduce a partition coefficient (and a consequent concentration jump) to describe the membrane functionality. Furthermore, mainly due to the fact that the model considers the coupling between the solute and the fluid balance (the Two Fluid part of TFEL), the description of the counter pressure is implicit: the counter pressure can be described as the direct effect of the solute-membrane interaction on the fluid flow. The momentum balance on the colloid phase describes the effect of the membrane on the transmission of the colloids through the membrane (selectivity) whereas, the momentum balance on the fluid predicts the flow resistance (counter pressure) due to osmosis. The calculation of the counter pressure then no longer relies on the semi-empirical law (derived from the Kedem-Katchalsky model) that assumes that the fluid flow is proportional to $\Delta p - \sigma \Delta \Pi$. It must be noted that the TFEL approach has similarities (and is physically consistent) with other model that describes the transfer inside a membrane with specific interactions, like the sorption-diffusion model [25]. The model can thus represent a way to unify the different approaches by considering that the membrane has a specific interaction with the solute (that can be linked to sorption, electrostatic interaction, partition) that can be generalized with the interfacial pressure. It could help to elucidate the "strange" transport mechanism of fluids at nanoscale[26] and therefore to progress by designing specific nanoscale molecule/pore interactions within artificial nano-pores in order to optimize the transport [27].

## 4.2 The interfacial pressure and the driving pressures

The energy landscape introduced in the TFEL model is scaled as a pressure (Fig. 2). The interfacial pressure, $\Pi_i$, can be compared to the fluid pressure, $p_f$, and to the osmotic pressure, $\Pi_{cc}$ ; these pressure terms represent the colloid-membrane, the fluid-fluid and the colloid-colloid interactions respectively. These pressures characterize the whole elastic (reversible) interaction energies and can be considered as a descriptor of the total energy storage in the mixture. The force per unit of volume resulting in pressure gradients are $\nabla p_f$, $\nabla \Pi_{cc}$ and $\phi \nabla \Pi_i$ for the fluid (fluid-fluid interactions), the colloid (colloid-colloid interactions) and for the interfacial interaction (colloid-interface interaction) respectively. The gradient of the interfacial pressure depends on the volume fraction of colloids, $\phi$ and the interactions force due to the interface, $d\Pi_i$. Pressure induced by the colloid-interface interaction can then be defined as the integral of the gradient, $\Pi_{ic} = \int \phi d\Pi_i$, that represents the total force exerted by the membrane on the colloids. This pressure contribution (referred to as exclusion pressure) is the term involved in the counter pressure (Eq. 15). The movement of the colloids and of the fluid occurs when these pressure contributions are released (the flow will occur in the direction of a negative gradient). As described with Eqs 10 and 11 and schematized in Fig. 8, the relative flow of the colloids in the mixture, $u_m - u_c$, is induced by the gradient of the colloid driving pressure, being the sum of the exclusion pressure, $\Pi_{ic}$, and the osmotic pressure, $\Pi_{cc}$. The gradient of the osmotic pressure leads to the diffusion of the colloids and is then responsible for the mass accumulation on membrane [28] and for the deposit reversibility [29], whereas the gradient of the exclusion pressure is responsible for the interaction induced migration of particles away from the membrane [13,30]. The velocity of the mixture, $u_m$, is due to a mixture driving pressure represented by the sum of the exclusion pressure and the fluid pressure. A gradient of fluid pressure leads to a forced fluid advection and thus to the permeation through the membrane. The gradient of the exclusion pressure leads to osmosis flow; the mechanism will be discussed in details in the following section.

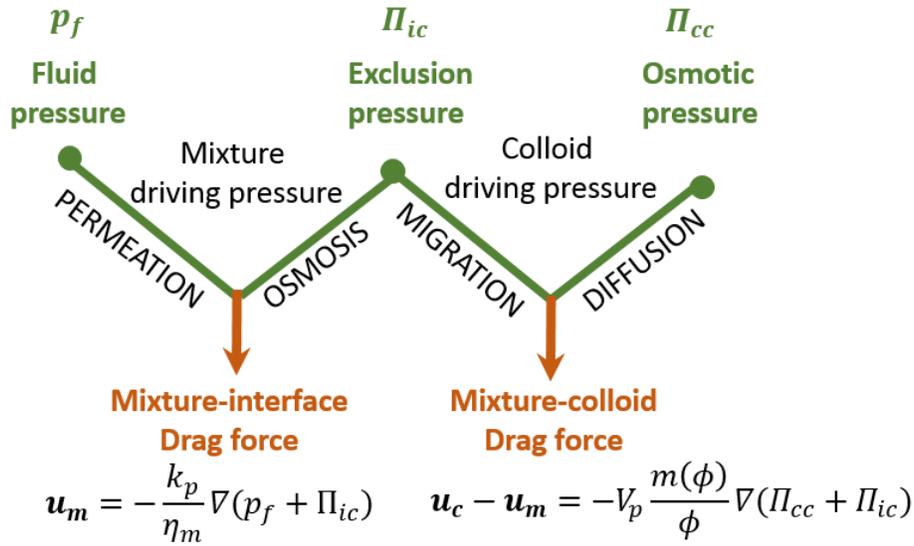

*Figure 8 : Schematic illustration of the role played by the pressure on the flow. The dissipative flows (orange arrows) are driven by the negative gradients of the elastic interaction energies related to the fluid pressure, the osmotic pressure and the interfacial pressure (green circles). The flow of the colloid phase with regard to the mixture, $u_m$-$u_c$, is driven by the colloid driving pressure (the sum of the exclusion and the osmotic pressures) that causes colloid migration and diffusion. The flow of the mixture, $u_m$, is driven by the mixture driving pressure (the sum of the exclusion and the fluid pressures) that causes osmosis and permeation.*

To illustrate these points quantitatively, Fig. 9 represents the variation of the colloid and the mixture driving pressure at the end of the simulation that describes the Abbé Nollet experiment (presented in section 3.2). As expected, the resulting colloid mixture pressure decreases along z to drive a flow of colloids from the concentrate to the diluate side (the membrane is not fully impermeable to the colloids). The mixture driving pressure increases along the membrane: a net osmotic flow exists from the diluate to the concentrate side. The gradient of the mixture driving pressure therefore represents the pressure drop needed to ensure the flow inside the membrane.

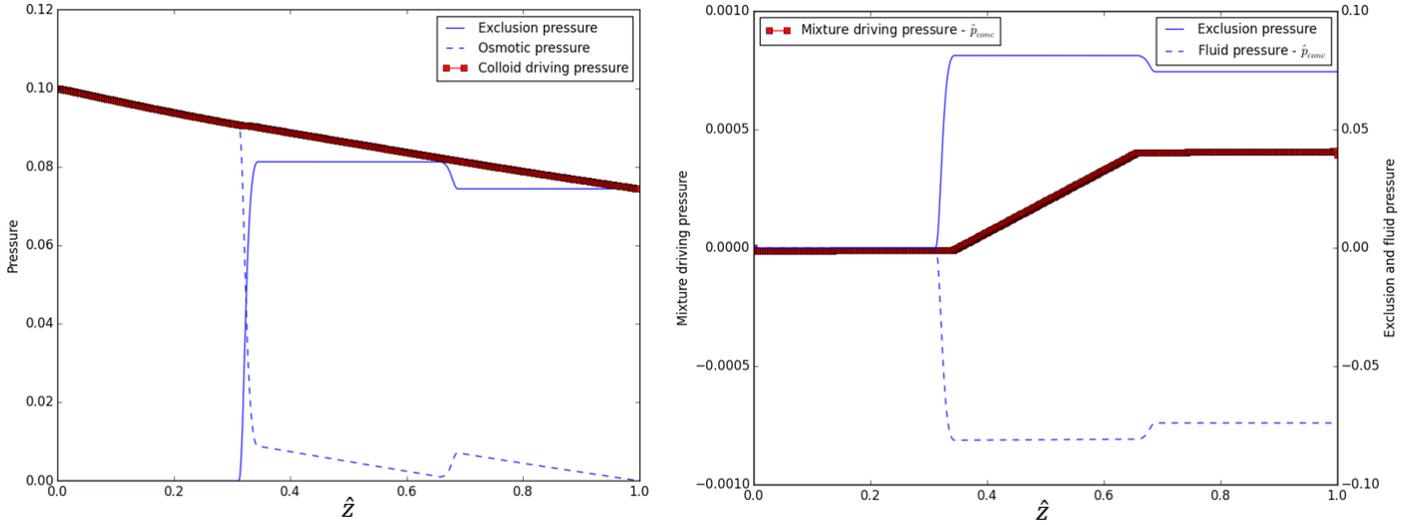

*Figure 9: The driving pressure for the colloid (on the left) and for the mixture (on the right). The driving pressure for colloids is the sum of the osmotic pressure and the exclusion pressure. The colloids are transferred from high to low colloid driving pressure (from left to right). The mixture driving pressure is the sum of the fluid pressure and the exclusion pressure. A mixture pressure gradient occurs inside the membrane where a pressure drop is needed to ensure the osmotic flow (from right to left).*

### 4.3 The exclusion pressure is responsible for a local fluid pressure change

The pressure of the fluid is linked, with Eq. 19, to the force exerted by the membrane on the solute. For the simulation that describes the Abbé Nollet experiment, the resulting fluid pressure as a function of the distance normal to the membrane surface is presented on Fig. 10. At the start of the experiment, the fluid pressures are equal in the two compartments and there is no flow rate. When the solute arrives at the membrane, by diffusion, the exclusion by the membrane on the concentrate side, $\phi \nabla \Pi_i > 0$, induces a decrease in fluid pressure, $\nabla p < 0$ as stated by Eq. 19 when there is no flow, $u_m = 0$:

$$-\nabla p - \phi \nabla \Pi_i = 0 \qquad (27)$$

Consequently, the pressure of the fluid at the interface, $p_{int}$, is smaller than that in the bulk, $p_{bulk}$; the difference being the exclusion pressure $\Pi_{ic}$:

$$p_{int} = p_{bulk} - \int \phi d\Pi_i = \Pi_{ic} \qquad (28)$$

The decrease in pressure near the membrane in the concentrate side leads to the pumping of the liquid from the dilute side. It should be noted that the pressure exerted on the membrane, $p_{int} + \int \phi d\Pi_i$, is still equal to the bulk pressure (as it will be measured with a sensor). For longer time, the pressure drop increases with the number of solute molecules accumulated in the exclusion layer (because of $\phi$ in the term $\phi \nabla \Pi_i$) leading to an increase in the osmosis flow rate (Fig. 7b). Similarly, a fluid pressure drop is observed in the dilute side as soon as the solute arrives in the dilute side (because the simulations are performed with a totally non-retentive membrane). However, the volume fraction being higher in the concentrate side, the net exclusion pressure (the sum of the exclusion pressure in the concentrate and diluate exclusion layers) leads to a net flux towards the concentrate side. At the end, the exclusion pressure, $\Pi_{ic}$, represents the complex interaction interplay between the fluid, the colloids and the membrane having for physical meaning: the pressure (a local depression) exerted on the fluid due to the colloid-interface interactions. This local pressure induced by the colloid concentration could also be at the origin of the flows associated with the Marangoni effect[31] or with the capillary osmosis[32] or the diffusiophoresis[33].

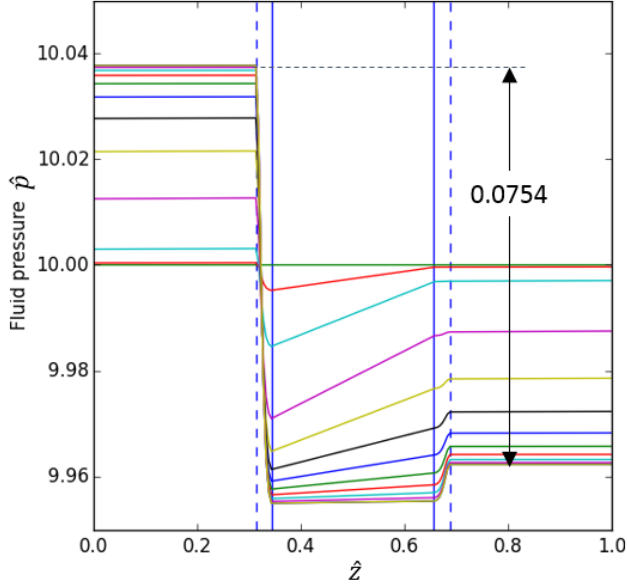

*Figure 10 : The dynamic variations of the dimensionless fluid pressure, $\hat{p}$, through the membrane during osmosis. The results correspond to the simulation and to the legend presented in Fig. 6. The exclusion of the colloids at the membrane interface leads to a local change in fluid pressure in the exclusion layers that initiates the osmotic flow.*

## 4.4 The solute-membrane interaction at the origin of the osmotic flow

The mechanism for the osmotic flow can be considered as the result of the interactions between the particles and the membrane as previously discussed by van't Hoff, Fermi and Einstein and recently reviewed [7,8]. From a force balance approach (schematized in Fig. 11), the scenario for the osmosis mechanism consists of the following steps:

1) In the exclusion layer, when particles arrive close to the membrane (due to the drag force induced by the flow, $nF_{drag}$, or due to the diffusion in a concentration gradient, $-\nabla \Pi_{cc}$), particles experience interactions with the membrane with a force, $-\phi \nabla \Pi_i$.
2) When the particles are arrested close to the membrane, the interaction forces counterbalance the forces, $nF_{drag} - \nabla \Pi_{cc}$
3) According to Newton's third law, these forces lead to equal and opposite reaction force on the fluid, $\nabla \Pi_{cc} - nF_{drag}$ directed away from the membrane (Fig. 9)
4) This force, $\nabla \Pi_{cc} - nF_{drag}$, participate to the displacement of the fluid away from the membrane. By considering the force balance on the colloid phase (Eq. 3), these forces are equal to the interactions force between the colloids and the membrane, $\nabla \Pi_{cc} - nF_{drag} = -\phi \nabla \Pi_i$.
5) These forces lead to a local fluid pressure drop in the exclusion layer, $\nabla p = -nF_{drag} + \nabla \Pi_{cc}$, when considering the momentum balance on the fluid (Eq. 28 and Fig. 10).
6) The pressure drop initiates the osmosis flow by ensuring the pumping of the liquid through the membrane.

The term osmosis initially introduced in 1854 by Thomas Graham finds here a particular meaning. Thomas Graham introduced this term from the ancient grec ὠσμός (ôsmos) which means push or pulse. In the mechanism described above, osmosis is directly related to the "push" exerted by the colloids on the semi-permeable membrane. Each time a colloid enters in the exclusion layer, the membrane pushes the colloid and, by reaction, this action leads to pushing the liquid away from the membrane. The fluid depression that is created by this movement initiates the osmosis. Thus, even if the membrane pushes only the colloids, a net force directed away from the membrane acts on the mixture and pushes the liquid away. If there is a difference in colloid concentration in the exclusion layer at either side of the membrane, the unbalanced force acting on the fluid will drive a net osmotic flow through the membrane.

In a simplified way, one can consider that the motor for osmosis is the osmotic pressure difference (it is why osmosis occurs) but the driving belt is the colloid /membrane interaction (it is how osmosis takes place). The TFEL model describes how the different forces play a role on the osmosis thanks to i) the consideration of finite exclusion layers with the Energy Landscape ii) the reciprocal contribution of the forces on the fluid and the colloid phase with the Two-Fluid approach. The model opens up interesting perspectives for the understanding of how the interaction landscape in exclusion layers impact on the dynamic of osmotic flows.

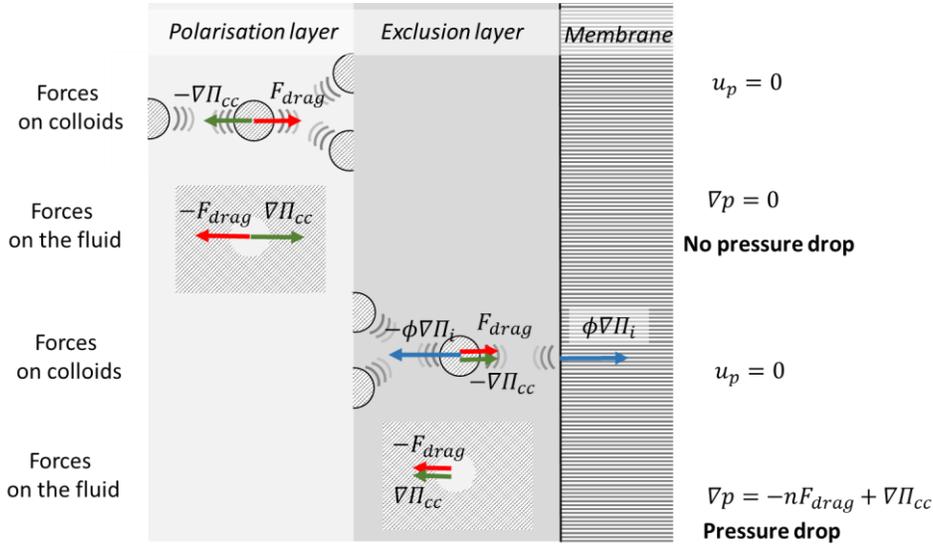

*Figure 11: Diagram of the main forces acting on the colloids and on the fluid in the polarization layer (first column) and in the exclusion layer (second column) at the membrane surface. In the polarization layer, the reaction force to the drag force acting on the fluid is counterbalanced by the force due to the motion of the solvent molecules (due to the gradient of solvent chemical potential). Consequently, there is no pressure drop [34]. On the contrary, in the exclusion layer, the forces acting on the fluid are not compensated thus leading to a pressure drop that is at the origin of the counter osmotic flow during filtration and the origin of the osmotic flow when the driving pressure is stopped.*

A questionable point is why the fluid pressure might be impacted by solute-membrane interactions: the local fluid pressure decreases when solutes are excluded because of the interaction with the membrane (as presented in Fig. 10 and discussed in the previous sections). As displayed in Fig. 11, inside the exclusion layer, the exclusion force acting on the colloids is not counterbalanced inside the fluid; the reaction of the exclusion interaction applies to the membrane. Newton's third law (action=reaction) is then violated for the mixture phase. This violation and the resulting breaking of the force symmetry can occur when there is a relative motion of interacting particles versus an interface [35]. In the exclusion layer, it leads (Fig. 11) to a net force on the fluid that is responsible for the counter osmotic flow during filtration and/or for an osmotic flow when the driving pressure is stopped. This exclusion pressure at the membrane surface goes with a local reduction in the fluid pressure, in order to keep constant the total pressure applied on the membrane surface: the loss in fluid pressure is equated to the pressure due to the normal force exerted by colloids on the membrane (the force violating the third newton law for the mixture). It must be noted, however, that Newton's third law holds for the complete "mixture-plus-interface" system: the net force on the fluid, $-\phi \nabla \Pi_i$, is counterbalanced by the force acting on the interface.

This local fluid pressure variation for colloidal dispersion near a wall has not yet been experimentally or theoretically demonstrated. However, the diffusion of colloids near a wall is still under investigation. The anisotropic near-wall hindered diffusion of particles [36] is currently studied for different conditions and for different interaction ranges with the wall [37]. The conclusion of these works is that the electrostatic force creates a need for an additional term to Einstein's theory of diffusion: the diffusion is frustrated and is no longer isotropic near a repulsive wall. The Stokes-Einstein diffusion theory assuming free or unhindered Brownian motion is no longer valid as particles approach a solid surface. Furthermore, the self-diffusion normal to the wall should vanish at long time scale (similar behavior has been shown for concentrated interacting colloids [38]). This anisotropic frustrated diffusion of colloids should, in turn, change the isotropic character of the solvent diffusion. On a microscopic point of view, the diffusion anisotropy and the loss of the kinetic energy of particles (due to the reduction of self-diffusion) should lead to a local decrease in the solvent kinetic energy (due to the molecular collisions) located between the "arrested" particles and the wall. In a simplified way, the arrested (or the frustrated diffusion) particles play the role of a screen or a damper for the solvent molecular collisions. The attenuation of these collisions should lead to a local decrease in fluid pressure (loss of collisions toward the membrane). However, it should be noted that it could be very difficult to measure such a phenomena: the membrane (or a pressure sensor including a membrane) still receives the fluid pressure of the bulk (the attenuated fluid pressure at the interface plus the exclusion pressure).

## 4.5 Pressure drop or not a pressure drop in the polarization layer?

In the context of the filtration with reverse-osmosis, nanofiltration or ultrafiltration, the transport processes within polarized layers have been theoretically and experimentally studied over the last decades. The pressure distribution along the polarized layers has been widely discussed but is still subject to debate. Several groups are considering, from theoretical analysis, that the pressure gradient should be zero in the polarization layers during ultrafiltration [34,39]. On an experimental point of view, pressure drop has been measured [40] while another authors [41] reports a constant pressure. More recently the various pressures involved in the polarization layers and their definition have been discussed in order to progress in the understanding [42]. These authors consider from an irreversible thermodynamic approach that the mixture thermodynamic pressure remains constant whereas the pervadic pressure (the pressure due to the fluid that is measured through a membrane) could change; this statement is close to the ones discussed in the previous section.

In this present paper, the two-fluid model enables accounting for the momentum balances schematized in Fig. 11. From this balance, there is no pressure drop in the polarization layer. The thermodynamic force and the drag force act in an opposite way both for the colloids phase (colloids are arrested in the polarization layer) and for the fluid (the friction on the fluid is compensated by the osmotic flow due to the concentration gradient): Newton's third law is not violated in the polarization layer. Inside the polarization layer (Fig. 9), the force applied on the fluid due to the friction, -$F_{drag}$, is counterbalanced by the force due to the motion of the solvent molecules, because of the chemical potential gradient in water. As a result, there is no pressure drop as already concluded by many authors [34,39]. No pressure drop (and therefore no dissipation) in the polarization layer is rather counter-intuitive when considering the picture of a slipping velocity on arrested particles. This picture has to be broken for two reasons: particles are Brownian (not fully arrested) and are not isolated, but rather in a positive concentration gradient. From a mechanical point of view, the thermodynamic force acting on the fluid (due to the difference in water activity generated by diffusive particles in a concentration gradient) equilibrates the frictional force on the fluid. Overall, the frictional energy entirely contributes to the concentration (unmixing) of particles: this internal exchange between fluid and particle phases does not generate a net dissipation (nor a pressure drop). The fluid pressure is thus constant, $\nabla p_f = 0$, in the polarization layer. This paper can provide an explanation to the historical debate about the pressure drop in polarized layers and resolve this controversy by considering:

- there is no pressure drop if only interactions between Brownian objects occur (in the colloidal dispersion or for a solution in the bulk and at a sufficient distance from the membrane) : Newton's third law (action=reaction) is applied for the mixture phase
- there is a pressure drop if interactions with non-Brownian objects take place (in the case of aggregated or gel phases or when close enough to a membrane to feel the interaction with the non-Brownian membrane): Newton's third law (action=reaction) is violated for the mixture phase.

This important difference is unraveled in the model by considering a polarization layer (where only interactions between Brownian objects take place) and an exclusion layer (where interactions with the membrane takes place). Such differences are also interesting to imagine and investigate more efficient separation processes. For example, separation inside a Brownian system (cage to cage diffusion in a colloidal Wigner glass [43]) or with an elastic membrane (storing the particle-membrane interaction) should be very efficient from an energetic point of view: the energy due to the separation (the relative motion between the particles and the fluid inherent to the separation) could be stored and then possibly further restituted.

# 5 Conclusions

A model describing the flow of a colloidal dispersion through a membrane has been developed with a two-fluid approach (to account for the colloid and the fluid transport) on an energy landscape (to represent the interfacial barrier). The model takes into account the interaction between the colloids and the membrane via an interfacial pressure. These interactions allow the description of the membrane selectivity for the colloids. Moreover, the colloid-membrane interactions enable, with a totally new way, the mechanism of osmotic flow to be described. Indeed, the colloid-membrane interactions lead to a local decrease in the fluid pressure near the membrane that is initiating the osmosis. The osmosis description thus no longer relies on the semi-empirical law derived from the Kedem-Katchalsky model. It has been demonstrated that the model is able to describe the dynamic of the osmosis flow through a membrane in the presence of a concentration gradient. It helps to understand the mechanical origin of the osmosis and to predict the dynamic variation of operating conditions during osmosis or reverse-osmosis processes. Furthermore, the model creates interesting possibilities to describe the dynamic of osmosis flow through biological membranes exhibiting specific energy map shapes.

# 6 Nomenclature

| | | |
|---|---|---|
| CP | counter pressure | (Pa) |
| D | diffusion coefficient | (m$^2$. s$^{-1}$) |
| F drag | drag force | (N) |
| Ex | exclusion number | (-) |
| k | Boltzmann constant | (J/K) |
| kp | permeability coefficient | (m$^2$) |
| K | Partition coefficient | (-) |
| Lp | membrane permeability | (m) |
| m | colloid mobility | (m$^2$. s$^{-1}$.J$^{-1}$) |
| n | number of colloidal particles | (m$^{-3}$) |
| p | pressure | (Pa) |
| Pe | Péclet number | (-) |
| T | temperature | (K) |
| Tr | colloid transmission | (-) |
| u | velocity | (m.s$^{-1}$) |
| Vp | volume of colloidal particles | (m$^{-3}$) |
| z | distance across the membrane | (m) |
| $\delta$ | thickness | (m) |
| $\phi$ | volume fraction | (-) |
| $\Phi$ | partition coefficient | (-) |
| $\eta$ | viscosity | (Pa.s) |
| $\Pi$ | osmotic, interfacial pressure | (Pa) |
| $\sigma$ | Staverman coefficient | (-) |

*Subscript :*

| | |
|---|---|
| c | particle |
| cc | colloid-colloid |
| ci | colloid-interface |
| f | fluid |
| m | mixture |
| BL | Boundary Layer |
| EX | Exclusion Layer |
| MB | Membrane Layer |

# 7 Acknowledgements

The author thanks Yannick Hallez, Fabien Chauvet, Leo Garcia, Martine Meireles and Pierre Aimar for the fruitful discussions on the model.

# 8 References

[1] L. Bocquet, J.-L. Barrat, Flow boundary conditions from nano- to micro-scales, Soft Matter. 3 (2007) 685–693. doi:10.1039/B616490K.

[2] O. Kedem, A. Katchalsky, Thermodynamic analysis of the permeability of biological membranes to non-electrolytes, Biochim. Biophys. Acta. 27 (1958) 229–246. doi:10.1016/0006-3002(58)90330-5.

[3] K.S. Spiegler, O. Kedem, Thermodynamics of hyperfiltration (reverse osmosis): criteria for efficient membranes, Desalination. 1 (1966) 311–326. doi:10.1016/S0011-9164(00)80018-1.

[4] J.H. van't Hoff, The role of osmotic pressure in the analogy between solutions and gases, J. Membr. Sci. 100 (1995) 39–44. doi:10.1016/0376-7388(94)00232-N.

[5] A. Einstein, Investigations on the theory of the Brownian movement, Ann Phys. (1905). http://www.physik.fu-

berlin.de/~kleinert/files/eins_brownian.pdf (accessed June 20, 2016).
[6] E. Fermi, Thermodynamics, Dover publication, 1936. http://store.doverpublications.com/048660361x.html (accessed June 20, 2016).
[7] S.S.S. Cardoso, J.H.E. Cartwright, Dynamics of osmosis in a porous medium, R. Soc. Open Sci. 1 (2014) 140352. doi:10.1098/rsos.140352.
[8] U. Lachish, Osmosis and thermodynamics, Am. J. Phys. 75 (2007) 997–998. doi:10.1119/1.2752822.
[9] P.H. Nelson, Osmosis and thermodynamics explained by solute blocking, Eur. Biophys. J. 46 (2017) 59–64. doi:10.1007/s00249-016-1137-y.
[10] E.M. Kramer, D.R. Myers, Osmosis is not driven by water dilution, Trends Plant Sci. 18 (2013) 195–197. doi:10.1016/j.tplants.2012.12.001.
[11] P. Bacchin, An energy map model for colloid transport, Chem. Eng. Sci. 158 (2017) 208–215. doi:10.1016/j.ces.2016.10.024.
[12] D. Wales, Energy Landscapes, Cambridge University Press, Cambridge, 2004. http://ebooks.cambridge.org/ref/id/CBO9780511721724 (accessed July 7, 2016).
[13] P. Bacchin, P. Aimar, V. Sanchez, Model for colloidal fouling of membranes, AIChE J. 41 (1995) 368–376. doi:10.1002/aic.690410218.
[14] J. Israelachvili, H. Wennerström, Role of hydration and water structure in biological and colloidal interactions, Nature. 379 (1996) 219–225. doi:10.1038/379219a0.
[15] E.J.W. Verwey, J.T.G. Overbeek, Theory of the Stability of Lyophobic Colloids, Courier Dover Publications, 1948.
[16] S. Sastry, P.G. Debenedetti, F.H. Stillinger, Signatures of distinct dynamical regimes in the energy landscape of a glass-forming liquid, Nature. 393 (1998) 554–557. doi:10.1038/31189.
[17] P.R. Nott, E. Guazzelli, O. Pouliquen, The suspension balance model revisited, Phys. Fluids 1994-Present. 23 (2011) 43304. doi:10.1063/1.3570921.
[18] H.M. Vollebregt, R.G.M. van der Sman, R.M. Boom, Suspension flow modelling in particle migration and microfiltration, Soft Matter. 6 (2010) 6052. doi:10.1039/c0sm00217h.
[19] G.C. Agbangla, P. Bacchin, E. Climent, Collective dynamics of flowing colloids during pore clogging, Soft Matter. 10 (2014) 6303–6315. doi:10.1039/C4SM00869C.
[20] G.S. Manning, Binary Diffusion and Bulk Flow through a Potential-Energy Profile: A Kinetic Basis for the Thermodynamic Equations of Flow through Membranes, J. Chem. Phys. 49 (1968) 2668–2675. doi:10.1063/1.1670468.
[21] A.J. Staverman, The theory of measurement of osmotic pressure, Recl. Trav. Chim. Pays-Bas. 70 (1951) 344–352. doi:10.1002/recl.19510700409.
[22] H.Y. Elmoazzen, J.A.W. Elliott, L.E. McGann, Osmotic Transport across Cell Membranes in Nondilute Solutions: A New Nondilute Solute Transport Equation, Biophys. J. 96 (2009) 2559–2571. doi:10.1016/j.bpj.2008.12.3929.
[23] J.E. Guyer, D. Wheeler, J.A. Warren, FiPy: Partial Differential Equations with Python, Comput. Sci. Eng. 11 (2009) 6–15. doi:10.1109/MCSE.2009.52.
[24] Abbé Nollet, Recherche sur les causes du bouillonnement des liquides, in: Hist. Académie R. Sci., Chez Gabriel Martin, Jean-Baptiste Coignard, Fils, Hippolyte-Louis Guerin, 1752: p. 57.
[25] J. Wang, D.S. Dlamini, A.K. Mishra, M.T.M. Pendergast, M.C.Y. Wong, B.B. Mamba, V. Freger, A.R.D. Verliefde, E.M.V. Hoek, A critical review of transport through osmotic membranes, J. Membr. Sci. 454 (2014) 516–537. doi:10.1016/j.memsci.2013.12.034.
[26] L. Bocquet, E. Charlaix, Nanofluidics, from bulk to interfaces, Chem. Soc. Rev. 39 (2010) 1073–1095. doi:10.1039/B909366B.
[27] S. Gravelle, L. Joly, F. Detcheverry, C. Ybert, C. Cottin-Bizonne, L. Bocquet, Optimizing water permeability through the hourglass shape of aquaporins, Proc. Natl. Acad. Sci. 110 (2013) 16367–16372. doi:10.1073/pnas.1306447110.
[28] Y. Bessiere, D.F. Fletcher, P. Bacchin, Numerical simulation of colloid dead-end filtration: Effect of membrane characteristics and operating conditions on matter accumulation, J. Membr. Sci. 313 (2008) 52–59. doi:10.1016/j.memsci.2007.12.064.
[29] B. Espinasse, P. Bacchin, P. Aimar, Filtration method characterizing the reversibility of colloidal fouling layers at a membrane surface: Analysis through critical flux and osmotic pressure, J. Colloid Interface Sci. 320 (2008) 483–490. doi:10.1016/j.jcis.2008.01.023.
[30] P. Bacchin, A. Marty, P. Duru, M. Meireles, P. Aimar, Colloidal surface interactions and membrane fouling: Investigations at pore scale, Adv. Colloid Interface Sci. 164 (2011) 2–11. doi:10.1016/j.cis.2010.10.005.
[31] J.L. Anderson, M.E. Lowell, D.C. Prieve, Motion of a particle generated by chemical gradients Part 1. Non-electrolytes, J. Fluid Mech. 117 (1982) 107–121. doi:10.1017/S0022112082001542.
[32] B.V. Derjaguin, S.S. Dukhin, M.M. Koptelova, Capillary osmosis through porous partitions and properties of boundary layers of solutions, J. Colloid Interface Sci. 38 (1972) 584–595. doi:10.1016/0021-9797(72)90392-X.
[33] D. Velegol, A. Garg, R. Guha, A. Kar, M. Kumar, Origins of concentration gradients for diffusiophoresis, Soft Matter. 12 (2016) 4686–4703. doi:10.1039/C6SM00052E.
[34] M. Elimelech, S. Bhattacharjee, A novel approach for modeling concentration polarization in crossflow membrane filtration based on the equivalence of osmotic pressure model and filtration theory, J. Membr. Sci. 145 (1998) 223–241.

[35] A.V. Ivlev, J. Bartnick, M. Heinen, C.-R. Du, V. Nosenko, H. Löwen, Statistical Mechanics where Newton's Third Law is Broken, Phys. Rev. X. 5 (2015) 11035. doi:10.1103/PhysRevX.5.011035.

[36] P. Huang, K.S. Breuer, Direct measurement of anisotropic near-wall hindered diffusion using total internal reflection velocimetry, Phys. Rev. E. 76 (2007) 46307. doi:10.1103/PhysRevE.76.046307.

[37] C.K. Choi, C.H. Margraves, K.D. Kihm, Examination of near-wall hindered Brownian diffusion of nanoparticles: Experimental comparison to theories by Brenner (1961) and Goldman et al. (1967), Phys. Fluids. 19 (2007) 103305. doi:10.1063/1.2798811.

[38] A. van Blaaderen, J. Peetermans, G. Maret, J.K.G. Dhont, Long-time self-diffusion of spherical colloidal particles measured with fluorescence recovery after photobleaching, J. Chem. Phys. 96 (1992) 4591–4603. doi:10.1063/1.462795.

[39] J.G. Wijmans, S. Nakao, J.W.A. Van Den Berg, F.R. Troelstra, C.A. Smolders, Hydrodynamic resistance of concentration polarization boundary layers in ultrafiltration, J. Membr. Sci. 22 (1985) 117–135. doi:10.1016/S0376-7388(00)80534-7.

[40] W. Zhang, C.R. Ethier, Direct pressure measurements in a hyaluronan ultrafiltration concentration polarization layer, Colloids Surf. Physicochem. Eng. Asp. 180 (2001) 63–73. doi:10.1016/S0927-7757(00)00755-X.

[41] A. Kim, C.H. Wang, M. Johnson, R. Kamm, The specific hydraulic conductivity of bovine serum albumin, Biorheology. 28 (1991) 401–419.

[42] S.S.L. Peppin, J. a. W. Elliott, M.G. Worster, Pressure and relative motion in colloidal suspensions, Phys. Fluids 1994-Present. 17 (2005) 53301. doi:10.1063/1.1915027.

[43] D. Bonn, H. Tanaka, G. Wegdam, H. Kellay, J. Meunier, Aging of a colloidal "Wigner" glass, EPL Europhys. Lett. 45 (1999) 52. doi:10.1209/epl/i1999-00130-3.